\documentclass[english,sort,aps,prd,reprint,nobibnotes,notitlepage,nofootinbib,showpacs,preprintnumbers]{revtex4-1}
\usepackage[latin9]{inputenc}
\usepackage{geometry}
\usepackage{fancyhdr}
\usepackage{color}
\usepackage{babel}
\usepackage{amsmath}
\usepackage{amssymb}
\usepackage{graphicx}
\usepackage{esint}
\usepackage[unicode=true,
 bookmarks=true,bookmarksnumbered=false,bookmarksopen=false,
 breaklinks=false,pdfborder={0 0 1},backref=false,colorlinks=true]
 {hyperref}
\hypersetup{pdftitle={Composite Inflation in the light of 2015 Planck data},
 pdfauthor={Phongpichit Channuie},
 citecolor=blue,linkcolor=blue,urlcolor=blue,citecolor=blue,linkcolor=blue}
\usepackage{breakurl}
\usepackage{multirow}
\usepackage{tikz}
\def\beq{\begin{equation}}
\def\eeq{\end{equation}}
\def\be{\begin{equation}}
\def\ee{\end{equation}}
\def\bea{\begin{eqnarray}}
\def\eea{\end{eqnarray}}
\def\ba{\begin{eqnarray}}
\def\ea{\end{eqnarray}}

\renewcommand{\(}{\left(}
\renewcommand{\)}{\right)}
\renewcommand{\[}{\left[}
\renewcommand{\]}{\right]}

\newcommand*\rfrac[2]{{}^{#1}\!/_{#2}}

\makeatletter
 \@ifundefined{textcolor}{}
 {%
   \definecolor{BLACK}{gray}{0}
   \definecolor{WHITE}{gray}{1}
   \definecolor{RED}{rgb}{1,0,0}
   \definecolor{GREEN}{rgb}{0,1,0}
   \definecolor{BLUE}{rgb}{0,0,1}
   \definecolor{CYAN}{cmyk}{1,0,0,0}
   \definecolor{MAGENTA}{cmyk}{0,1,0,0}
   \definecolor{YELLOW}{cmyk}{0,0,1,0}
 }

\usepackage{babel}
\@ifundefined{definecolor}{color}{}
\usepackage{babel}

\usepackage{fancyhdr}\@ifundefined{definecolor}{color}{}
\usepackage{babel}

\usepackage{xcolor}
\usepackage{fancyheadings}
\makeatother

\begin{document}

\title{Composite Inflation in the light of 2015 Planck data}

\author{Phongpichit Channuie}

\email{channuie@gmail.com}

\date{\today}

\affiliation{School of Science, Walailak University, Thasala, \\Nakhon Si Thammarat, 80160, Thailand}


\begin{abstract}
In this work, we examine cosmological constraints on models of composite inflation based on the slow-roll approximation by using the recent Planck measurement. We compare the spectral index of curvature perturbation (and its running) and the tensor-to-scalar ratio predicted by such models with Planck 2015 data. We find that the predictions of technicolor inflation are nicely consistent with the Planck analysis. Moreover, the predictions from the second model, glueball inflation, are in good agreement with the Planck data at $2\sigma$C.L. However, the final two models, super glueball inflation and orientifold inflation, favor only the rather large value of the tensor-to-scalar ratio of which the predictions are in tension with the Planck analysis. 
\end{abstract}
\maketitle

\section{Introduction}

It was widely believed that there was a period of accelerating expansion in the very early universe traditionally known as inflation. The inflationary paradigm \cite{Alex,KSa,KSa1,DKa,GUT} tends to solve important issues that plagued the standard big bang theory and successfully describes the generation and evolution of the observed large-scale structures of the universe. The inflationary scenario is formulated so far by the introduction of (elementary) scalar fields (called inflaton) with a nearly flat potential \cite{new,new1,chaotic,natural,natural1,Linde}. 

However, we can imagine the possibility that the inflaton need not be an elementary degree of freedom. The authors in \cite{Channuie:2011rq,Bezrukov:2011mv,Channuie:2012bv} have shown that it is possible to construct models in which the inflaton emerges as a composite state of a four-dimensional strongly coupled theory.  We called these types of models composite inflation (see for a recent review \cite{Channuie:2014ysa}). There was another interesting model of composite inflation based on holographic approach \cite{Evans:2010tf}.

The precise measurements recently released by Planck \cite{Ade:2015lrj} of the Cosmic Microwave Background (CMB) anisotropies covering the entire sky and over a broad range of scales provide a powerful probe of inflationary cosmology. At the inflationary frontiers, many cosmological paradigms, e.g. $R^{2}$ inflation \cite{Starobinsky:1980te}, power-law inflation \cite{Linde:1983gd}, $\alpha$-attractors \cite{Kallosh:2013yoa}, and many others \cite{Martin:2013tda}, are being challenged by current observations.  

In this work, we will examine the constraints on models of composite inflation based on the slow-roll approximation and compare the spectral index of curvature perturbation (and its running) and the tensor-to-scalar ratio predicted by such models with Planck 2015 data. We also compare our results with those of some selected inflationary models in the last section.

\section{Strong dynamics non-minimally coupled to gravity}
We start in this section by writing the general action for composite inflation in the Jordan frame (JF) taking the form for the scalar-tensor theory of gravity \cite{Fuji} as
\begin{eqnarray}
\mathcal{S}_{\rm JF}=\int d^{4}x \sqrt{-g}\Big[- \frac{M^{2}_{\rm P}+\xi\varphi^{\rfrac{2}{d}}}{2}R +{\cal L}(\varphi) \Big], \label{action}
\end{eqnarray}
where ${\cal L}(\varphi)$ is the low-energy effective Lagrangian for the field $\varphi$ which has mass dimension $d$ and the non-minimal coupling to gravity is characterised by the dimensionless coupling $\xi$. In this framework, the non-analytic power of the field $\varphi$ emerges if one requires a dimensionless coupling with the Ricci scalar, $R$. In our case, the effective Lagrangian ${\cal L}(\varphi)$ can be in general written as
\begin{eqnarray}
{\cal L}(\varphi) = g^{\mu\nu}\varphi^{\rfrac{(2-2d)}{d}}\partial_{\mu}\varphi\partial_{\nu}\varphi - V(\varphi), \label{lagrange}
\end{eqnarray}
with $V(\varphi)$ the potential of the underlying theories. In this investigation, we will analyse the dynamics in the Einstein frame (EF). To this end, we diagonalize the gravity-scalar sector in Eq.(\ref{action}) via the conformal transformation
\begin{eqnarray}
g_{\mu\nu} \rightarrow \tilde{g}_{\mu\nu} = \Omega(\varphi)^{2} g_{\mu\nu},\,\,\Omega(\varphi)^{2} = \frac{M^{2}_{\rm P}+\xi\varphi^{\rfrac{2}{d}}}{M^{2}_{\rm P}}. \label{xform}
\end{eqnarray}
Performing the conformal transformation yields the Einstein frame action and we find the resulting action written in terms of the canonically normalized field $\chi$ as
\begin{eqnarray}
\mathcal{S}_{\rm EF}=\int d^{4}x \sqrt{-g}\Big[- \frac{M^{2}_{\rm P}}{2}R +\frac{1}{2}g^{\mu\nu}\partial_{\mu}\chi\partial_{\nu}\chi - U(\chi) \Big], \label{action1}
\end{eqnarray}
where we have removed the tildes for convenience and the potential $U(\chi)$ is given by
\begin{eqnarray}
U(\chi) = \Omega^{-4}V(\varphi). \label{Upot}
\end{eqnarray}
Here the canonically normalized field $\chi$ is related to the original one $\varphi$ via
\begin{eqnarray}
\frac{d\chi}{d\varphi}= \[2\Omega^{-2}\left(1+\frac{3\xi^{2}}{d^{2}M^{2}_{\rm P}}\Omega^{-2}\varphi^{\rfrac{2}{d}}\right)\varphi^{\rfrac{(2-2d)}{d}}\]^{\rfrac{1}{2}}. \label{normfield}
\end{eqnarray}
In the context of single-field inflation, the slow-roll parameters in terms of $U$ and $\chi$ are defined by
\begin{eqnarray}
\epsilon & = & \frac{M^{2}_{\rm P}}{2}\left(\frac{dU/d\chi}{U}\right)^{2} = \frac{M^{2}_{\rm P}}{2}\left(\frac{U'}{U}\frac{1}{\chi'}\right)^{2}, \cr
\eta & = & M^{2}_{\rm P}\left(\frac{d^{2}U/d\chi^{2}}{U}\right) = M^{2}_{\rm P}\left(\frac{U'}{\chi'}\right)'\left(\frac{1}{U\chi'}\right), \cr
\zeta & = & M^{4}_{\rm P}\frac{dU}{d\chi}\left(\frac{d^{3}U/d\chi^{3}}{U^{2}}\right) = M^{2}_{\rm P}\left(\frac{U'}{\chi'^{2}U^{2}}\right)\[\left(\frac{U'}{\chi'}\right)'\frac{1}{\chi'}\]', \cr
N & = & \frac{1}{M^{2}_{\rm P}}\int^{\chi_{\rm ini}}_{\chi_{\rm end}}\frac{U}{dU/d\chi}d\chi = \frac{1}{M^{2}_{\rm P}}\int^{\varphi_{\rm ini}}_{\varphi_{\rm end}}\left(\frac{U}{U'}\right)\chi'^{2}d\varphi\,, \label{paras}
\end{eqnarray}
where primes denote differentiation with respect to the field $\varphi$ and $M_{P}$ is the reduced Planck mass. However, it is cumbersome to obtain an explicit solution of Eq.(\ref{normfield}). As a result, we will instead express parameters in terms of the field $\varphi$. In this work, we will compare our predictions with the experimental results via the relative strength of the tensor perturbation, i.e. the tensor-to-scalar ratio $r$ and the spectral index of curvature perturbation $n_{s}$ and its running $n'_{s}$. In terms of the slow-roll parameters, these observables are given by \cite{Lyth:2009}
\begin{align}
r & \simeq 16\epsilon, \label{para2}\\
n_{s} &\simeq  1 - 6\epsilon + 2\eta\,, \label{para1}\\
n'_{s} & = \frac{dn_{s}}{d\ln k} \simeq -24\epsilon^{2} +16\epsilon\eta -2\zeta\,, \label{para0}
\end{align}
where all parameters given above are evaluated at the field value $\chi_{\rm ini.}$ or $\varphi_{\rm ini.}$. Moreover, in order to generate the proper amplitude of the density fluctuations, the potential must satisfy at $\sigma_{\rm WMAP}$ the normalization condition such that $U/\epsilon \simeq (0.0276\,M_{\rm P})^{4}$ \cite{Bezrukov:2008ut} corresponding to the initial value of the inflaton fields. The constraints on $r$ imply the upper limit on the energy scale of inflation via \cite{Ade:2015lrj}
\begin{align}
U_{*}=\frac{3\pi^{2}}{2}A_{s}rM^{4}_{\rm P}=(1.88\times 10^{16}\,\,{\rm GeV})^{4}\frac{r}{0.10} \,, \label{scale}
\end{align}
where $A_{s}$ is the scalar power spectrum amplitude given by $U/(24\pi^{2}M^{4}_{\rm P}\epsilon)$. It has been shown that the scale of composite inflation, for all the models presented here, is of the order of the grand unified energy scale. In order to obtain the maximum number of allowed e-foldings for specific models of inflation, we need the following relation \cite{Ade:2015lrj}:
\begin{eqnarray}
N_{*} \approx  67 &-& \ln\left(\frac{k_{*}}{a_{0}H_{0}}\right) +\frac{1}{4}\ln\left(\frac{U_{*}}{M^{4}_{\rm P}}\right)+\frac{1}{4}\ln\left(\frac{U_{*}}{\rho_{\rm end}}\right)\nonumber\\& - &\frac{1-3w_{\rm int}}{12(1+w_{\rm int})}\ln\left(\frac{\rho_{\rm th}}{\rho_{\rm end}}\right) - \frac{1}{12}\ln(g_{\rm th}) \,, \label{emax}
\end{eqnarray}
where the universe has thermalized at the energy scale $\rho_{\rm th}$, $\rho_{\rm end}$ is the energy density at the end of inflation, $a_{0}H_{0}$ is the present Hubble radius, $U_{*}$ is the potential energy when $k_{*}$ left the Hubble radius during inflation, $w_{\rm int}$ denotes the effective equation of state between the energy scale specified by $\rho_{\rm in}$ and the end of inflation, and $g_{\rm th}$ is the number of effective bossing degrees of freedom at the energy scale $\rho_{\rm th}$.

The final four terms of Eq.(\ref{emax}) characterizes the uncertainty in the various energy scales connected with inflation. In typical models of inflation, these factors are not expected to be too large; whilst the first two terms of Eq.(\ref{emax}) are model independent, with the second term being roughly 5 for $k_{*}=0.05$\,Mpc$^{-1}$ \cite{Planck:2013jfk}. The magnitude of the last term is negligible since it gives a shift of only $0.58$ for the extreme value $g_{\rm th}=10^{3}$.

\section{Contact with observations}
The Planck 2015 data \cite{Ade:2015lrj} recently released reported the spectral index of curvature perturbations to be $n_{s} = 0.968 \pm 0.006$ and tightly constrained its scale dependence to $dn_{s}/d \ln k = -0.003 \pm 0.007$ when combined with the Planck lensing likelihood. The upper bound on the tensor-to-scalar ratio is $r < 0.11$ (95$\%$\,CL) which is consistent with the B-mode polarization constraint $r < 0.12$ (95 $\%$\,CL) obtained from a joint analysis of the BICEP2/$Keck\,Array$ and $Planck$ data \cite{Ade:2015tva}. Regarding the inflationary machinery given in the previous section, we will examine below the composite models of inflation.
\begin{figure*}[t]
\begin{center}
 \includegraphics[width=0.49\linewidth]{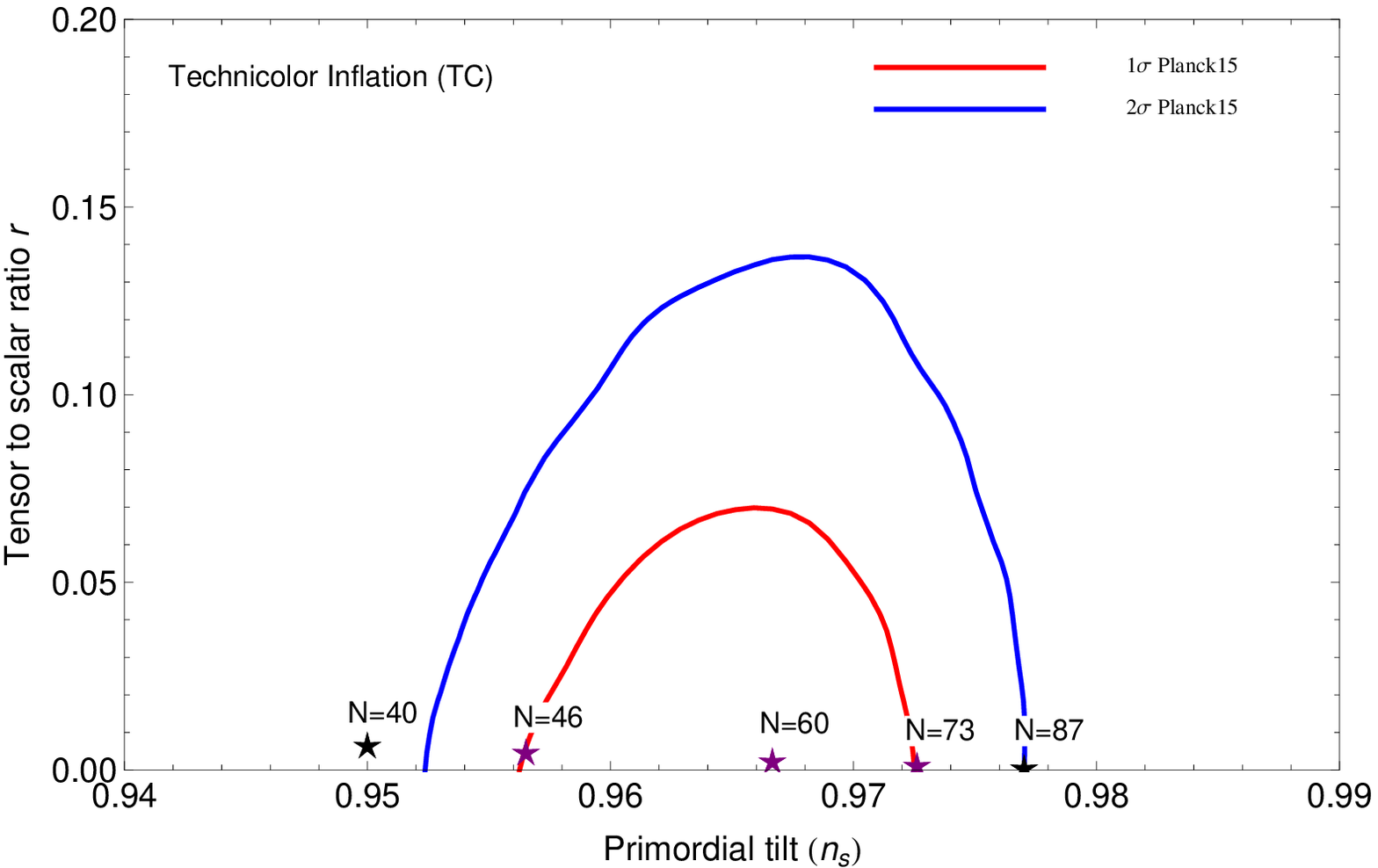}
 \includegraphics[width=0.49\linewidth]{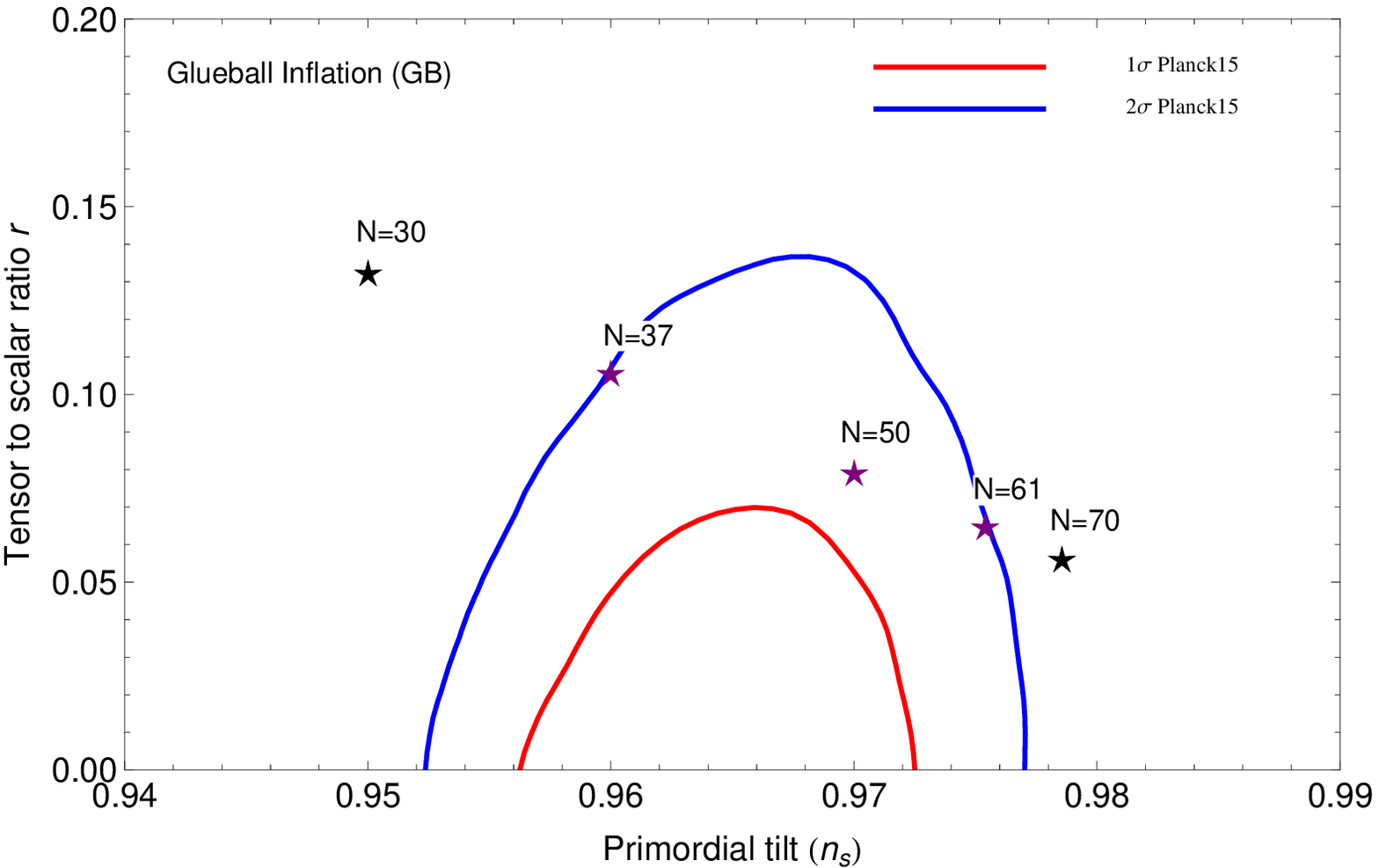}
  \includegraphics[width=0.49\linewidth]{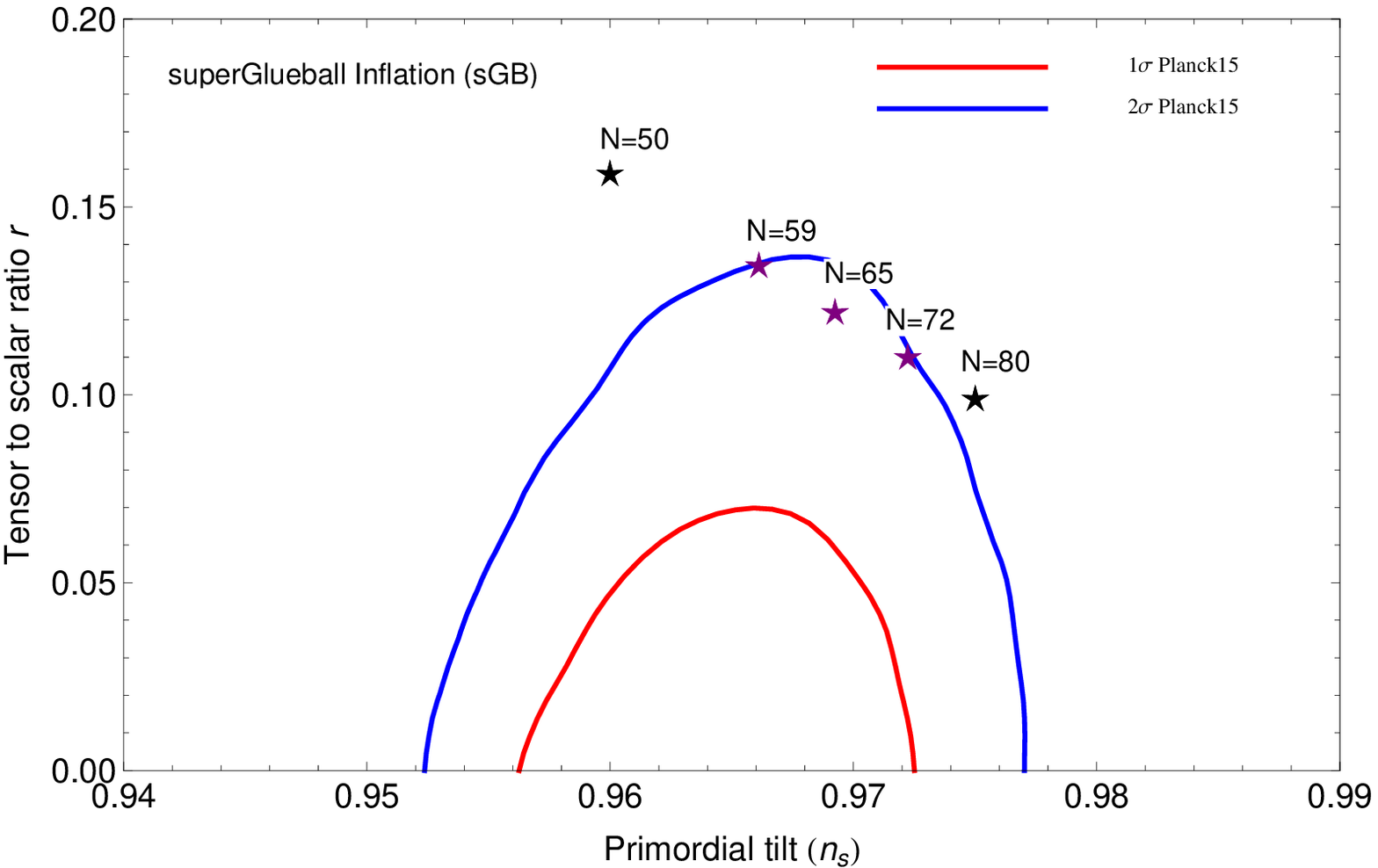}
   \includegraphics[width=0.49\linewidth]{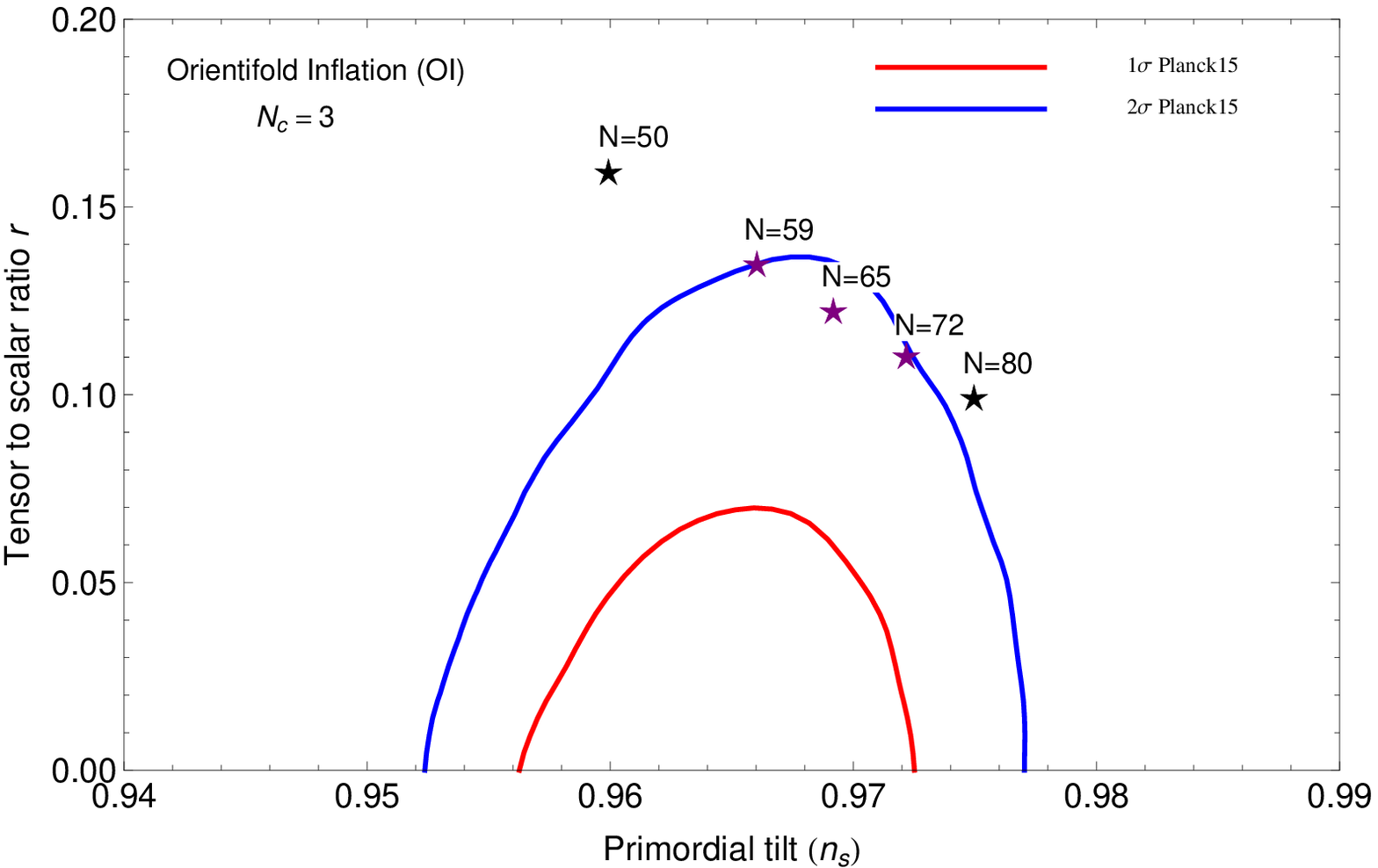}
 \caption{The theoretical predictions in the $(r-n_{s})$ plane for different models of composite inflation with Planck$'15$ results for TT, TE, EE, +lowP and assuming $\Lambda$CDM + r \cite{Ade:2015lrj}.}
 \label{fig2}
\end{center}
\end{figure*}

\vspace{4 mm}
%%%%%%%%%%%%%%%%%%%%%%%
{\it Technicolor (TC) inflation}
%%%%%%%%%%%%%%%%%%%%%%%
\vspace{3 mm}

\begin{figure*}[t]
\begin{center}
 \includegraphics[width=0.49\linewidth]{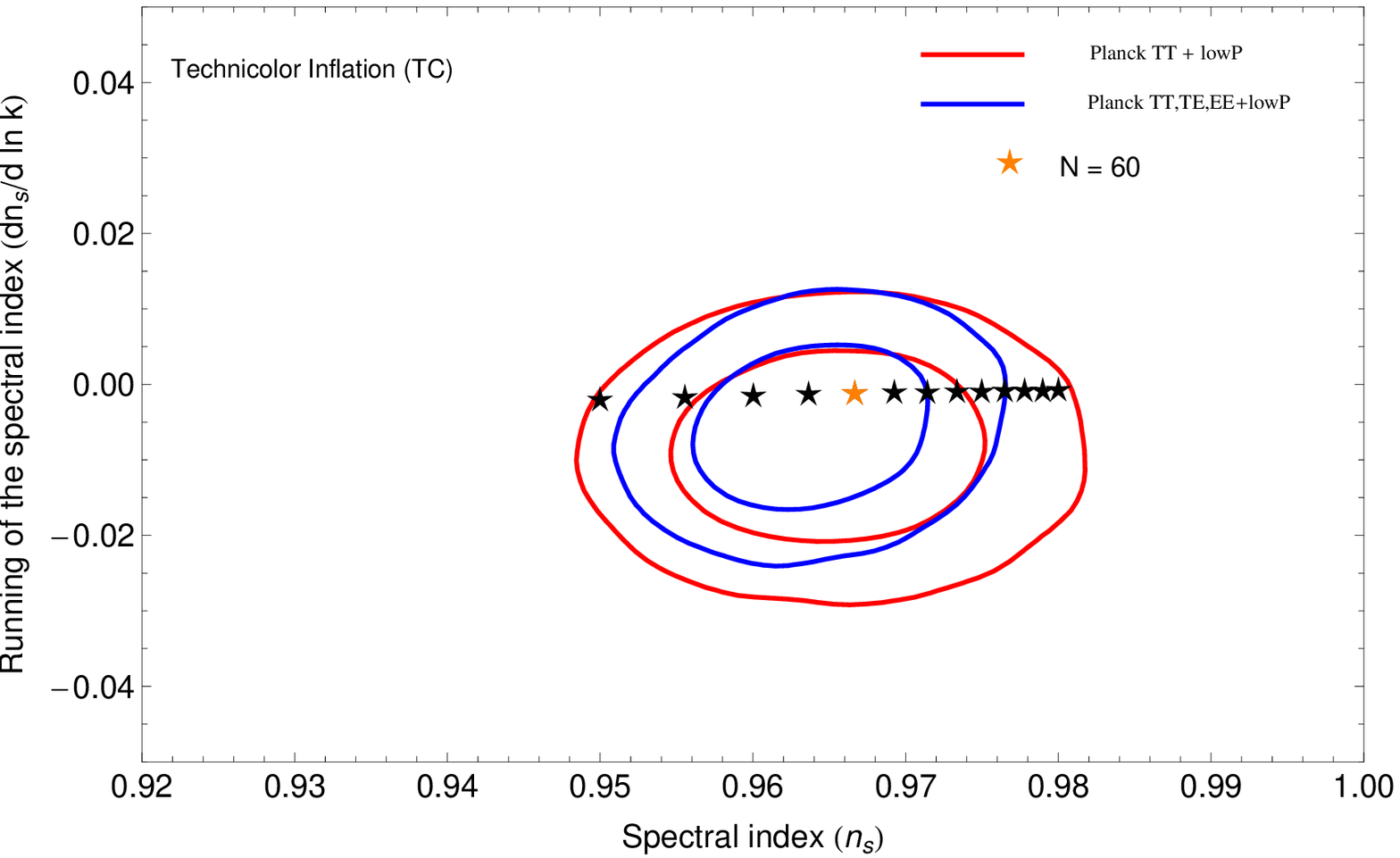}
 \includegraphics[width=0.49\linewidth]{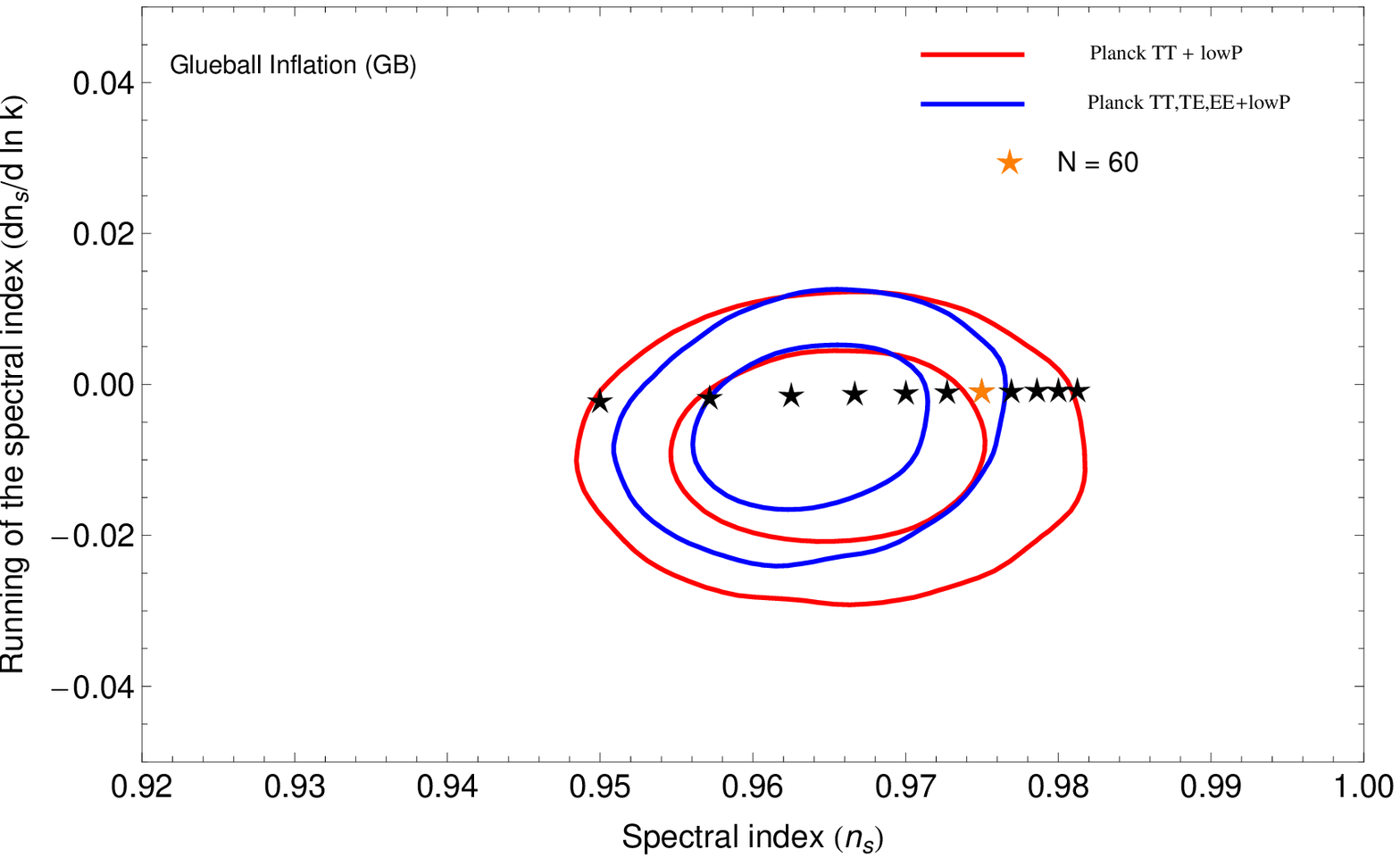}
 \includegraphics[width=0.49\linewidth]{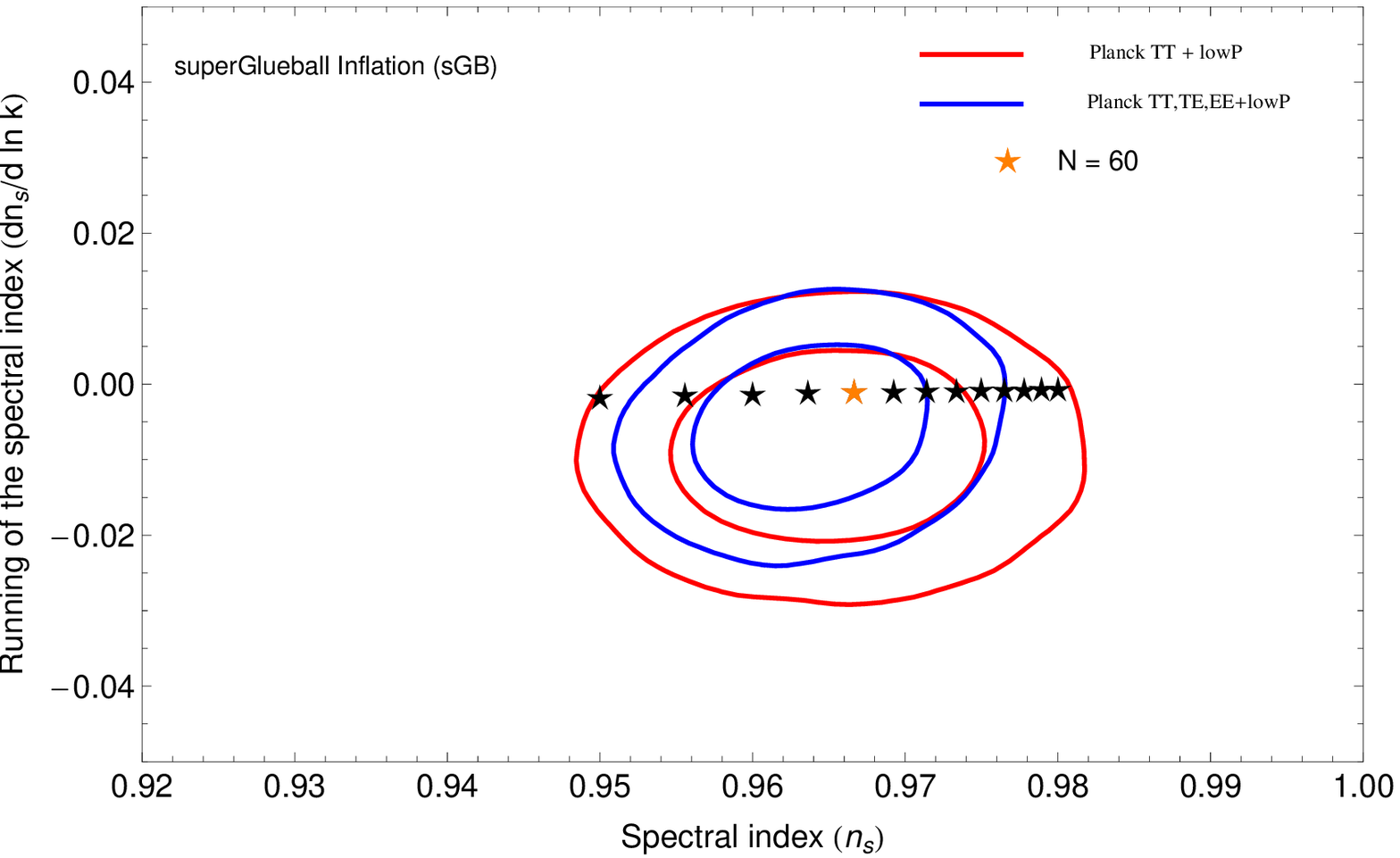}
 \includegraphics[width=0.49\linewidth]{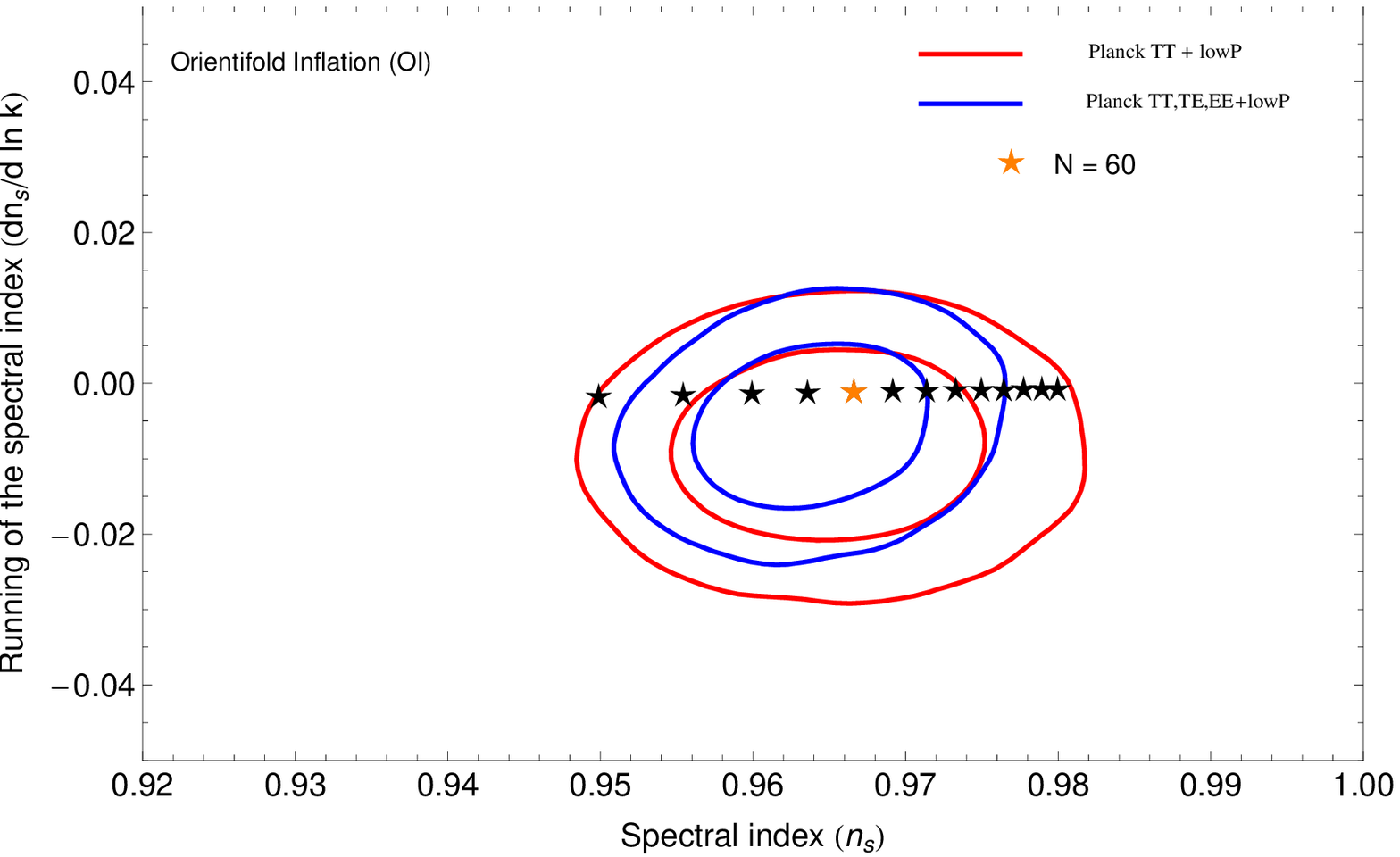}
 \caption{Marginalized joint $68\%$ and $95\%$\,C.L. for ($n_{s},\,dn_{s}/d ln k$) using Planck TT+lowP and Planck TT,TE,EE+lowP \cite{Ade:2015lrj}. For comparison, the stars show the prediction for different composite models.}
 \label{fig4}
\end{center}
\end{figure*}

In this scenario, we engaged the simplest models of technicolor known as the minimal walking technicolor (MWT) theory with the standard (slow-roll) inflationary paradigm as a template for composite inflation and name it here, in short, the TC model. In this case, we have $d=1$. The action for this model is given in \cite{Channuie:2011rq} with the potential in the Einstein frame:
\begin{eqnarray}
{\cal U}_{\rm TC}(\varphi) \simeq \frac{\kappa M^{4}_{P}}{4\xi^{2}}\left(1+\frac{M^{2}_{\rm P}}{\xi\varphi^{2}}\right)^{-2}\,,
\label{u-tc}
\end{eqnarray}
with $\kappa$ the inflaton self-coupling. Now inserting Eq.(\ref{u-tc}) into the slow-roll parameters we obtain for the large field approximation
\begin{eqnarray}
\epsilon  \simeq \frac{4M^{4}_{\rm P}}{3\xi^{2}\varphi^{4}}, \,\,\,\,\eta \simeq -\frac{4M^{2}_{\rm P}}{3\xi\varphi^{2}}, \,\,\,\,\zeta  \simeq  \frac{16M^{4}_{\rm P}}{9\xi^{2}\varphi^{4}}.
\end{eqnarray}
At the end of inflation, i.e. $\epsilon(\varphi_{\rm end}) =1$, we find $\varphi_{\rm end} \sim M_{\rm P}/\xi$. The number of e-foldings in this case reads 
\begin{eqnarray}
N \simeq \frac{6}{8M^{2}_{\rm P}/\xi}\Big[\varphi^{2}_{\rm ini}-\varphi^{2}_{\rm end}\Big]\quad{\rm with}\quad \varphi_{\rm ini}\gg\varphi_{\rm end}\,.\label{foldtc}
\end{eqnarray}
Regarding the inflaton field at the onset of inflation, we obtain $\varphi \simeq \sqrt{8N/6}M_{\rm P}/\xi \sim 9M_{\rm P}/\xi$ for $N=60$ e-foldings. It is straightforward to express the inflationary predictions in terms of the number of e-foldings, and we find to the lowest order in the slow-roll approximation: $n_{s}\simeq1-2/N,\,\,n'_{s}\simeq -2/N^{2}-12/N^{3},\,\,r\simeq 12/N^2$. Now let us briefly present our results. In Fig.(\ref{fig2}), we show the confidence contours in the ($n_{s},\,r$) plane. On the top-left panel, the predictions are based on the potential (\ref{u-tc}) which are in excellent agreement with the data provided the Planck team. Specifically, we obtain for this model $n_{s}=0.96667,\,r=0.00333$ which lie well inside the Planck data for $N=60$ at $1\sigma$ region of the contours. The normalization of the amplitude of the fluctuations implies that the scale of this composite model is of the order of the grand unified energy scale. Concerning the running spectral index, we obtain $n'_{s}=-0.00061$ for $N=60$ e-foldings. From Fig.(\ref{fig4}), we discover for the TC model that the running of the scalar spectral index does not significantly change as a function of $n_{s}$. More specifically for the TC model, the third term of Eq.(\ref{emax}) is roughly -5 and the fourth term is slightly greater than $O(1)$. Hence we reasonably obtain for this model the maximum number of e-foldings $N_{*} \approx 68$. 

\vspace{4 mm}
%%%%%%%%%%%%%%%%%%%%%%%
{\it pure Yang-Mills or Glueball (GB) inflation}
%%%%%%%%%%%%%%%%%%%%%%%
\vspace{3 mm}

This model of composite inflation is driven by gluonic-type fields. In this case, the inflaton emerges as the interpolating field describing the lightest glueball associated to a pure Yang-Mills theory. We then engage this theory non-minimally to gravity lying the standard (slow-roll) inflationary paradigm as a template for composite inflation and name it here, in short, the GB model. In this case, we have $d=4$. This model is characterised by the following potential \cite{Bezrukov:2011mv} in the Einstein frame
\begin{eqnarray}
{\cal U}_{\rm GB}(\varphi) \simeq \frac{2M^{4}_{P}}{\xi^{2}}\ln\Big(\varphi/\Lambda\Big)\,.
\label{u-gb}
\end{eqnarray}
with $\Lambda$ the strongly coupled scale which is found to be the typical scale of grand unification, i.e. ${\cal O}(10^{16})$\,GeV. Now inserting Eq.(\ref{u-gb}) into the slow-roll parameters, we find for this model
\begin{eqnarray}
\epsilon  \simeq \frac{1}{12\ln\left(\varphi/\Lambda\right)^2}, \,\,\,\,\eta \simeq 0, \,\,\,\,\zeta  \simeq  0.
\end{eqnarray}
At the end of inflation, i.e. $\epsilon(\varphi_{\rm end}) =1$, we find for the model $\varphi_{\rm end}/\Lambda \simeq \exp(\sqrt{1/12}) \sim 1.3$. In the large field limits, the number of e-foldings reads
\begin{eqnarray}
N \simeq 3\Big(\ln\(\varphi_{\rm ini}/\Lambda\)^{2} - \ln\(\varphi_{\rm end}/\Lambda\)^{2}\Big)\,.\label{foldgb}
\end{eqnarray}
Regarding the inflaton field at the beginning of inflation, we obtain $\varphi_{\rm ini}/\Lambda \simeq \exp(\sqrt{N/3}) \sim 87$ for $N=60$ e-foldings. The inflationary predictions in terms of the number of e-foldings in the Einstein frame parameters to the lowest order in the slow-roll approximation in this case read $n_{s}\simeq 1-3/2N,\,\,n'_{s}\simeq -3/2N^{2},\,\,r\simeq 4/N$. In Fig.(\ref{fig2}), we place our predictions to the confidence contours in the ($n_{s},\,r$) plane. On the top-right panel of Fig.(\ref{fig2}), the predictions are based on the potential (\ref{u-gb}) which are good agreement with the data. 

Specifically, we obtain for this model $n_{s}=0.975,\,r=0.06667$ which lie inside the Planck data for $N=60$ at $2\sigma$ region of the contours. For this model, the constraint on $r$ encourages the scale of inflation which matches the one of the grand unified energy scale, like the first model. Concerning the running spectral index, we obtain $n'_{s}=-0.00042$ for $N=60$ e-foldings. From Fig.(\ref{fig4}), we find for this model that the running of the scalar spectral index does not significantly change as a function of $n_{s}$. More specifically for the GB model, the third term of Eq.(\ref{emax}) is roughly -5 and the fourth term is slightly less than $O(1)$. Hence we obtain the maximum number of allowed e-foldings for this model $N_{*} \approx 67$. 

\vspace{4 mm}
%%%%%%%%%%%%%%%%%%%%%%%
{\it super Glueball (sGB) inflation}
%%%%%%%%%%%%%%%%%%%%%%%
\vspace{3 mm}

In this scenario, the inflaton is designed to be the gluino-ball state in the super Yang-Mills theory. We have also engaged the underlying theory non-minimally to gravity and assumed standard (slow-roll) inflationary paradigm. We name it here, in short, the sGB model. In this case, we have $d=3$. The potential of this model in the Einstein frame takes the form \cite{Channuie:2012bv}
\begin{eqnarray}
{\cal U}_{\rm sGB}(\varphi) \simeq \frac{4\alpha}{N^{2}_{c}}\frac{M^{4}_{P}}{\xi^{2}}\ln\Big(\varphi/\Lambda\Big)^{2}\,, \label{u-sgb}
\end{eqnarray}
where $N_{c}$ is the number of colors and $\alpha$ is a constant which is given by the underlying theory and is expected to be of order unity \cite{Feo:2004mr}. Here $\Lambda$ is an invariant scale of the theory which is $N_{c}$-dependent. By increasing the number of underlying colors, it is possible to lower scale of inflation in this case \cite{Channuie:2012bv}. After inserting Eq.(\ref{u-sgb}) into Eq.(\ref{paras}), we obtain the slow-roll parameters in this model:
\begin{eqnarray}
\epsilon \simeq \frac{1}{3\ln\left(\varphi/\Lambda\right)^2}, \,\,\,\,\eta \simeq \frac{1}{3\ln\left(\varphi/\Lambda\right)^2}, \,\,\,\,\zeta  \simeq  0.
\end{eqnarray}
At the end of inflation, i.e. $\epsilon(\varphi_{\rm end}) =1$, we find for the model $\varphi_{\rm end}/\Lambda \simeq \exp(\sqrt{1/3}) \sim 1.8$. In the large field limits, the number of e-foldings is given by
\begin{eqnarray}
N \simeq \frac{3}{2}\Big(\ln\(\varphi_{\rm ini}/\Lambda\)^{2} - \ln\(\varphi_{\rm end}/\Lambda\)^{2}\Big)\,.\label{foldsgb}
\end{eqnarray}
Regarding the inflaton field at the beginning of inflation, we obtain $\varphi_{\rm ini}/\Lambda \simeq \exp(\sqrt{2N/3}) \sim 558$ for $N=60$ e-foldings. To the lowest order in the slow-roll approximation, the inflationary predictions in terms of the number of e-foldings in the Einstein frame parameters for this model are $n_{s}\simeq 1-2/N,\,\,n'_{s}\simeq -2/N^{2},\,\,r\simeq 8/N$. On the bottom-left panel, the predictions are based on the potential (\ref{u-sgb}) which are partly agreement with the data at $2\sigma$\,C.L. 

Specifically, we obtain for this model $n_{s}=0.96667,\,r=0.13333$ which lie at the boundary of the $2\sigma$ region of the contours for $N=60$ e-foldings. This model also encourages the scale of composite model is of the order of $10^{16}$ GeV, like the first-two models. Concerning the running spectral index, we obtain $n'_{s}=-0.00056$ for $N=60$ e-foldings. Apparently, from Fig.(\ref{fig4}), we find for this model that the running of the scalar spectral index does not significantly change as a function of $n_{s}$. More specifically for the sGB model, the third term of Eq.(\ref{emax}) is roughly -5 and the fourth term is $O(1)$. Hence we obtain for this model the reasonable maximum number of e-foldings $N_{*} \approx 67$. 

\vspace{4 mm}
%%%%%%%%%%%%%%%%%%%%%%%
{\it Orientifold (OI) inflation}
%%%%%%%%%%%%%%%%%%%%%%%
\vspace{3 mm}

In this work, the gluino field of supersymmetric gluodynamics is replaced by two Weyl fields which can be formed as one Dirac spinor. We have engage the underlying theory non-minimally to gravity and assumed the slow-roll approximation. In this case, we also have $d=3$. This model is characterised by the following potential \cite{Channuie:2012bv} in the Einstein frame 
\begin{eqnarray}
{\cal U}_{\rm OI}(\varphi) \simeq \frac{4\alpha}{N^{2}_{c}}\frac{M^{4}_{P}}{\xi^{2}}\left[\ln\Big(\varphi/\Lambda\Big)^{2} - \frac{\beta}{9}\right]\,,
\label{u-oi}
\end{eqnarray}
where $\alpha$ is a constant, $\beta={\cal O}(1/N_{c})$ and $\Lambda$ is the scale of the theory. Notice that at large $N_{c}$ limit, i.e. $\beta\rightarrow 0$, this theory maps into the preceding one. After inserting Eq.(\ref{u-oi}) into the slow-roll parameters, we obtain for the large field approximation
\begin{eqnarray}
\epsilon = \eta \simeq  \frac{1}{3\ln\left(\varphi/\Lambda\right)^2}\Big(1+\frac{2\beta}{9\ln\left(\varphi/\Lambda\right)^2}\Big),\,\zeta \simeq 0\,.
\end{eqnarray}
At the end of inflation, i.e. $\epsilon(\varphi_{\rm end}) =1$, we find for the model $\varphi_{\rm end}/\Lambda \simeq \exp(\sqrt{1/3})(1+\beta/3\sqrt{3}) \sim 1.8(1+0.19\beta)$. In the large field limits, the number of e-foldings is given by
\begin{eqnarray}
N \simeq \left(\frac{3}{2}\ln\left(\frac{\varphi}{\Lambda}\right)^{2}\Big[1-\frac{2\ln\ln\left(\frac{\varphi}{\Lambda}\right)}{81\ln\left(\frac{\varphi}{\Lambda}\right)^{2}}\beta\Big]\right)^{\varphi_{\rm ini}}_{\varphi_{\rm end}}\,.\label{foldsgb}
\end{eqnarray}
Regarding the inflaton field at the beginning of inflation, we obtain $\varphi_{\rm ini}/\Lambda \simeq \exp(\sqrt{2N/3})(1+[4+3\ln(2N)]\beta/12\sqrt{6N}) \sim 558(1+0.008\beta)$ for $N=60$ e-foldings. To the lowest order in the slow-roll approximation, the inflationary predictions in terms of the number of e-foldings in the Einstein frame parameters for this model are $n_{s}\simeq 1-2/N - 2\beta/3N^{2},\,\,n'_{s}\simeq -2/N^{2}+\ln(2N)\beta/3N^{3},\,\,r\simeq 8/N+8\beta/3N^{2}$. On the bottom-right panel of Fig.(\ref{fig2}), the predictions are based on the potential (\ref{u-sgb}) which are also agreement with the data provided the Planck satellite at $2\sigma$\,C.L. 

Specifically, we obtain for this model $n_{s}=0.96661,\,r=0.13358$ which lie at the boundary of the $2\sigma$ region of the contours for $N=60$ e-foldings and $N_{c}=3$. Concerning the running spectral index, we obtain $n'_{s}=-0.00055$ for $N=60$ e-foldings. For this model, we discover that the running of the scalar spectral index does not significantly change as a function of $n_{s}$ (see Fig.(\ref{fig4})). More specifically for the OI model, the third term of Eq.(\ref{emax}) is roughly -5 and the fourth term is also $O(1)$. Hence we obtain for this model the reasonable maximum number of e-foldings $N_{*} \approx 67$. 

\section{Discussion and Conclusion}

Let us summarize our investigation by comparing our results with some selected inflationary models. The first model we anticipate to compare with is $R^{2}$ inflation in which the corresponding potential in the Einstein frame takes the form: $V(\phi)=\Lambda^{4}(1-\exp(-\sqrt{2/3}\phi/M_{\rm P})^{2}$ \cite{Starobinsky:1980te}. This model predicts $n_{s}\simeq 1-2/N,\,r\simeq 12/N^{2}$. Apparently, it turns out that the predictions coincide with those of the TC model. Undoubtedly, at this level of the investigation we can not distinguish theses two models. In order to reconcile with this point, the (pre)reheating mechanism in the models will encourage us to typically identify the distinction. In the meantime, the details of studying preheating mechanism in the TC model with the minimalistic approach can be found in \cite{Channuie:2016xmq}. However, a thorough and comprehensive study of reheating mechanism is still required and we will leave this interesting topic for our future investigation.

The second model of inflation we are going to discuss is the chaotic inflation. In this scenario, the corresponding potential takes the power-law form $V(\phi)=\Lambda^{4}(\phi/M_{\rm P})^{n}$ \cite{Linde:1983gd}. Specifically, the predictions for the scalar spectral index and the tensor-to-scalar ratio at first order in the slow-roll approximation for $n=2$ are $n_{s} \simeq 1 - 2/N$ and $r \simeq 8/N$, respectively. Apparently, we find that the predictions of this model coincide with those of the sGB model. Although the underlying theories of these two models are completely different, the comprehensive study of reheating mechanism is still required in order to further distinguish them. 

There are other interesting models. Recently, two classes of inflationary models motivated by the developments in conformal symmetry and supergravity have been proposed called the $\alpha$-attractors. The first one is known as the E-models in which the corresponding potential in the Einstein frame takes the form \cite{Kallosh:2013yoa}: $V(\phi)=\Lambda^{4}(1 - \exp[-2\sqrt{2}\phi/(\sqrt{3\alpha}M_{\rm P})])^{2}$ with $\alpha$ a constant. To lowest order in the slow-roll approximation, this model predicts $n_{s} \simeq 1-8[1+\exp(\sqrt{2}\phi/(\sqrt{3\alpha}M_{\rm P}))]/[3\alpha(1-\exp(\sqrt{2}\phi/(\sqrt{3\alpha}M_{\rm P})))],\,\,r\simeq 64/[3\alpha(1-\exp(\sqrt{2}\phi/(\sqrt{3\alpha}M_{\rm P})))]$ \cite{Ade:2015lrj}. The second class of models is known as the T-models characterized by the following potential: $V(\phi)=\Lambda^{4}\tanh^{2m}(\phi/\sqrt{6\alpha}M_{\rm P})))$ \cite{Kallosh:2013yoa}. The slow-roll predictions for this potential can be found in \cite{Ade:2015lrj}. Yet, another model of inflation is called hyperbolic inflation in which the corresponding potential takes the hyperbolic form: $V(\phi)=A(\sinh[\sqrt{3}(\gamma-\gamma_{\phi})(\phi-\phi_{0})/\sqrt{\gamma_{\phi}}])$ where $A,\,\gamma,\,\gamma_{\phi}$ are constants. The slow-roll predictions for this model can be found in \cite{Basilakos:2015sza}.

Especially, however, the potential of the GB model is quite subtle since it becomes negative at some values of the field and its minimum is also negative. Regarding the corresponding potential, the ground state reads $\left<\varphi\right> = e^{-1/4}\Lambda$. Using this field value, we find the minimum of the potential becomes negative $V_{\rm GB,\,min} = -\Lambda^{4}/(2e)$. Moreover, at the inflationary level, the potential will be negative for the field value $\varphi<\Lambda$. Fortunately, we find for this model that inflation ends before its potential become negative ($\varphi_{\rm end} \approx 1.3 \Lambda$). In the context of cosmology with negative potentials, the authors of \cite{Felder:2002jk,Wang:2003as} investigated cosmological evolution in models in which the effective potentials become negative at some values of the inflaton field. Several qualitatively new features emerge from such models as compared to those of the positive ones.

In this present work, we have examined cosmological constraints on models of composite inflation based on the slow-roll approximation and compared the spectral index of curvature perturbation (and its running) and the tensor-to-scalar ratio predicted by such models with Planck 2015 data. We have also examined for each model the reasonable maximum number of allowed e-foldings, taking into account the scale of fluctuations at which the predictions are compared to Planck data. We discovered that the predictions of technicolor inflation are nicely consistent with the Planck analysis. Moreover, the predictions from glueball inflation are in good agreement with the Planck data at $2\sigma$\,C.L. However, the final two models, super glueball inflation and orientifold inflation, favor only the rather large value of the tensor-to-scalar ratio of which the predictions are in tension with cosmological measurements released by Planck. Hopefully, further improvement of the accuracy of these measurements may turn out to be critical in falsifying composite scenarios.

\paragraph*{\bf Acknowledgments}
The author thanks Fedor Bezrukov for illuminating suggestions. This work is financially supported by the Institute for the Promotion of Teaching Science and Technology (IPST) under the project of the \lq\lq Research Fund for DPST Graduate with First Placement\rq\rq\, under Grant No. 033/2557 and by the Thailand Research Fund (TRF) under the project of the \lq\lq TRF Grant for New Researcher\rq\rq\, with Grant No. TRG5780143.

\end{document}